\documentclass[preprint,prc,tightenlines,showpacs,superscriptaddress,nofootinbib]{revtex4-1}
\usepackage{amsfonts}
\usepackage{amssymb}
\usepackage{amsmath}
\usepackage{graphicx,color}
\usepackage{dcolumn}
\usepackage{epsfig}
\usepackage{bm}
\usepackage{bbm}
\usepackage{ulem}

\usepackage{color}

\begin{document}
\title{Pomeron-LQCD model of $J/\Psi$ photo-production on the nucleon
}
\author{T.-S. H. Lee}
\affiliation{Physics Division, Argonne National Laboratory, Argonne, Illinois 60439, USA}

\begin{abstract}

Based on the vector meson dominance assumption, a  Hamiltonian model has been developed
to investigate $J/\Psi$ photo-production reaction on the nucleon
by  using the $J/\Psi$-nucleon potential extracted from
a lattice QCD calculation of
 Phys. Rev. D{\bf 82}, 091501 (2010).  It is found that 
the predicted total cross sections are comparable to the recent data of
$J/\Psi$ photo-production reaction  from Jefferson Laboratory.
The model is then extended to include the two-gluon exchange amplitude
modeled by Donnachie and Lanshoff within Regge Phenomenology. The resulting Pomeron-LQCD model can
then explain the data up to invariant mass $W=$ 300 GeV.
Future improvements needed to reduce  the uncertainties of the predictions are discussed.
The need of  an accurate extraction of $J/\Psi$-N potential at short distances from LQCD is illustrated.

\end{abstract}



\maketitle

\section{Introduction}
 
The $J/\Psi$-nucleon ($N$) interaction is  mediated by gluon exchanges within Quantum Chromodynamics (QCD). 
It has been investigated  by using lattice QCD (LQCD) and the data of the extracted $J/\Psi$-N potential $v_{LQCD}(r)$
have been published 
by  Kawanai and Sasaki\cite{sasaki,sasaki-1}. The purpose of this work is to explore how this LQCD potential can be 
used to predict  $J/\Psi$ photo-production reaction cross sections, and how it can be combined
with the Pomeron-exchange model, as developed in Refs.\cite{dl,ohlee,wulee}, to explain  the recent
data\cite{jlab} from Jefferson Laboratory (JLab) and also the earlier data up to invariant mass $W=300$ GeV.

In section II, we  review the Pomeron-exchange model formulated in Refs.\cite{ohlee,wulee}.
A model based on the vector meson dominance (VMD) and the LQCD potential $v_{LQCD}(r)$ is
presented in section III. In section IV,  the model is extended to  include the amplitudes generated
from the 
 Pomeron-exchange model.
The discussions on necessary future improvements  are given in section V.

\section{Pomeron-exchange model}
We use the convention\cite{gw} that the plane-wave state, $|\vec{k}>$,  is normalized as  
$<\vec{k}|\vec{k}^{\,\,'}>=\delta(\vec{k}-\vec{k}^{\,\,'})$ and the S-matrix is related to the
scattering T-matrix
by $S_{fi} =\delta_{fi}- 2\pi\, i T_{fi}$.
In the center of mass frame, the 
differential cross section of vector meson ($V$) photo-production
reaction, $\gamma (\vec{q})+N(-\vec{q})\rightarrow V(\vec{k})+N(-\vec{k})$,
is  calculated from
\begin{eqnarray}
\frac{d\sigma_{VN,\gamma N}}{d\Omega}(W)
=\frac{(2\pi)^4}{q^2}\rho_{VN}(k)\rho_{\gamma N}(q)\frac{1}{4}
\sum_{\lambda_V,m'_s}\sum_{\lambda_\gamma,m_s}
|<\vec{k},\lambda_V,m'_s|T_{VN,\gamma N}(W)|\vec{q},\lambda_\gamma  m_s>|^2
\label{eq:crst-gnjn}
\end{eqnarray}
where $ \rho_{VN}(k)=\frac{kE_V(k)E_N(k)}{W}$ and $ \rho_{\gamma N}(q)=\frac{q^2E_N(q)}{W}$,
$m_s$ denotes the z-component of the nucleon spin, and $\lambda_V$ and $\lambda_\gamma$
are the helicities of  vector meson $V$ and photon $\gamma$, respectively.
The magnitudes of $k=|\vec{k}|$ and $q=|\vec{q}|$ are defined by the invariant mass
$W=q+E_N(q)=E_V(k)+E_N(k)$.
In the Pomeron-exchange model developed in Refs.\cite{ohlee,wulee},  the scattering amplitude is 
written as
\begin{eqnarray}
<\vec{k},\lambda_V,m'_s|T_{VN,\gamma N}(W)|\vec{q},\lambda_\gamma m_s>
=<k\lambda_{V};p_fm_s'|T_{P}|q_i\lambda_\gamma,p_i m_s>
\label{eq:t-pom}
\end{eqnarray}
where the four momenta are $k=(E_V(\vec{k}),\vec{k})$, $p_f=(E_N(\vec{k}), -\vec{k})$,
$q_i=(q, \vec{q})$, $p_i=(E_N(\vec{q}), -\vec{q})$, and 
\begin{eqnarray}
<k\lambda_{V};p_fm_s'|T_{P}|q\lambda_\gamma,p_i m_s>
& =&
\frac{1}{(2\pi)^3}\sqrt{\frac{ m_Nm_N }{4 E_{V}(\vec{k}) E_N(\vec{p}_f)
|\vec{q}|E_N(\vec{p}_i) }}
\epsilon_\nu(q,\lambda_\gamma)
[j^{\nu}_{\lambda_{V},\,m'_s,\,m_s}(k,p_f,q,p_i)]
\nonumber\\
\label{eq:pomt}
\end{eqnarray}
In the above equation,  $\epsilon_\nu(q,\lambda_\gamma)$ is
the polarization vector of photon. The current matrix element in Eq.(\ref{eq:pomt}) is

\begin{eqnarray}
j^{\nu}_{\lambda_{V},\,m'_s,\,m_s}(k,p_f,q,p_i)=\bar{u}(p_f,m'_s)
\epsilon^*_\mu(k,\lambda_{V})\mathcal{M}^{\mu\nu}_\mathbb{P}(k,p_f,q,p_i)
u(p_i,m_s)\,, \label{eq:ampjpsip}
\end{eqnarray}
where $u(p,m_s)$ is the nucleon spinor (with the normalization
$\bar{u}(p,m_s){u}(p,m'_s) = \delta_{m_s,m'_s}$) ,
$\epsilon_\nu(k,\lambda_V)$ is the polarization vector of vector meson $V$.
In the amplitude defined in Eq.(\ref{eq:ampjpsip}), $\mathcal{M}^{\mu\nu}(k,p_f,q,p_i)$, for the Pomeron-exchange
mechanism can be written as:
\begin{equation}
\mathcal{M}^{\mu\nu}_\mathbb{P}(k,p_f,q,p_i) = G_\mathbb{P}(s,t)
\mathcal{T}^{\mu\nu}_\mathbb{P}(k,p_f,q,p_i)
\label{eq:MP}
\end{equation}
with
\begin{eqnarray}
\mathcal{T}^{\mu\nu}_\mathbb{P}(q,p,q',p') = [i 12 \frac{eM_V^2}{f_V}]
 [\beta_{q_{V}}F_V(t)]][\beta_{u/d} F_1(t)] \{ q\!\!\!/ \, g^{\mu\nu} - q^\mu \gamma^\nu \}  \, ,
\label{eq:pom-a}
\end{eqnarray}
where for $V=\rho, \omega, \phi, J/\Psi, y(1s)$, the masses are
$M_V=775.50, 782.65, 1019.45, 3096.91, 9460.00$ MeV, and 
 $f_V= 5.3, 15.2, 13.4, 11.2, 40.53$ are determined from
the decay widths of $V\rightarrow e^+e^-$. The parameters
$\beta_{q_{V}}$ ($\beta_{u/d}$) defines the coupling of the Pomeron with the
quark $q_{V}$ ($u$ or $d$ )in the vector meson $V$ (nucleon $N$).
In Eq.(\ref{eq:pom-a}) we have also introduced
a form factor for the Pomeron-vector meson vertex  as
\begin{eqnarray}
F_V(t)=\frac{1}{M_V^2-t} \left( \frac{2\mu_0^2}{2\mu_0^2 + M_V^2 - t} \right)
\label{eq:f1v}
\end{eqnarray}
where $t=(q-k)^2=(p_f-p_i)^2$. By using the Pomeron-photon analogy\cite{dl} ,
the form factor for the Pomeron-nucleon vertex is defined by
the isoscalar electromagnetic form factor of the nucleon as
\begin{equation}
F_1(t) = \frac{4M_N^2 - 2.8 t}{(4M_N^2 - t)(1-t/0.71)^2}.
\label{eq:f1}
\end{equation}
Here $t$ is in unit of GeV$^2$, and $M_N$ is the proton mass.

The crucial ingredient of  Regge Phenomenology is the propagator $G_\mathbb{P}$
for the Pomeron in Eq. (\ref{eq:MP}).
It is of the following form :
\begin{equation}
G_\mathbb{P} = \left(\frac{s}{s_0}\right)^{\alpha_P(t)-1}
\exp\left\{ - \frac{i\pi}{2} \left[ \alpha_P(t)-1 \right]
\right\} \,,
\label{eq:regge-g}
\end{equation}
where $s=(q+p_i)^2=W^2$, $ \alpha_P (t) = \alpha_0 + \alpha'_P t$.

By fitting the data of $\rho^0$, $\omega$, $\phi$,
photo-production\cite{ohlee}, the parameters of the
model have been determined:
 $\mu_0=  1.1$ GeV$^2$, $\beta_{u/d}=2.07$ GeV$^{-1}$,
$\beta_{s}=1.38$ GeV$^{-1}$, $\alpha_0=1.08$ for $\rho$ and $\omega$,
$\alpha_0=1.12$ for $\phi$,
and $\alpha'_P = 1/s_0^{} = 0.25$ GeV$^{-2}$.
For the heavy quark systems, we find  
that with the 
same $\mu_0^2$, $\beta_{u/d}$, and $\alpha'_P $,
the $J/\Psi$ and $y(1s)$ photo-production  data can be fitted by setting
$\beta_c = 0.32$ GeV$^{-1}$ and $\beta_b = 0.45$ GeV$^{-1}$
and choosing a larger $\alpha_0=1.25$.

\begin{figure}[h] \vspace{+2.0cm}
\begin{center}
\includegraphics[width=0.6\columnwidth,angle=0]{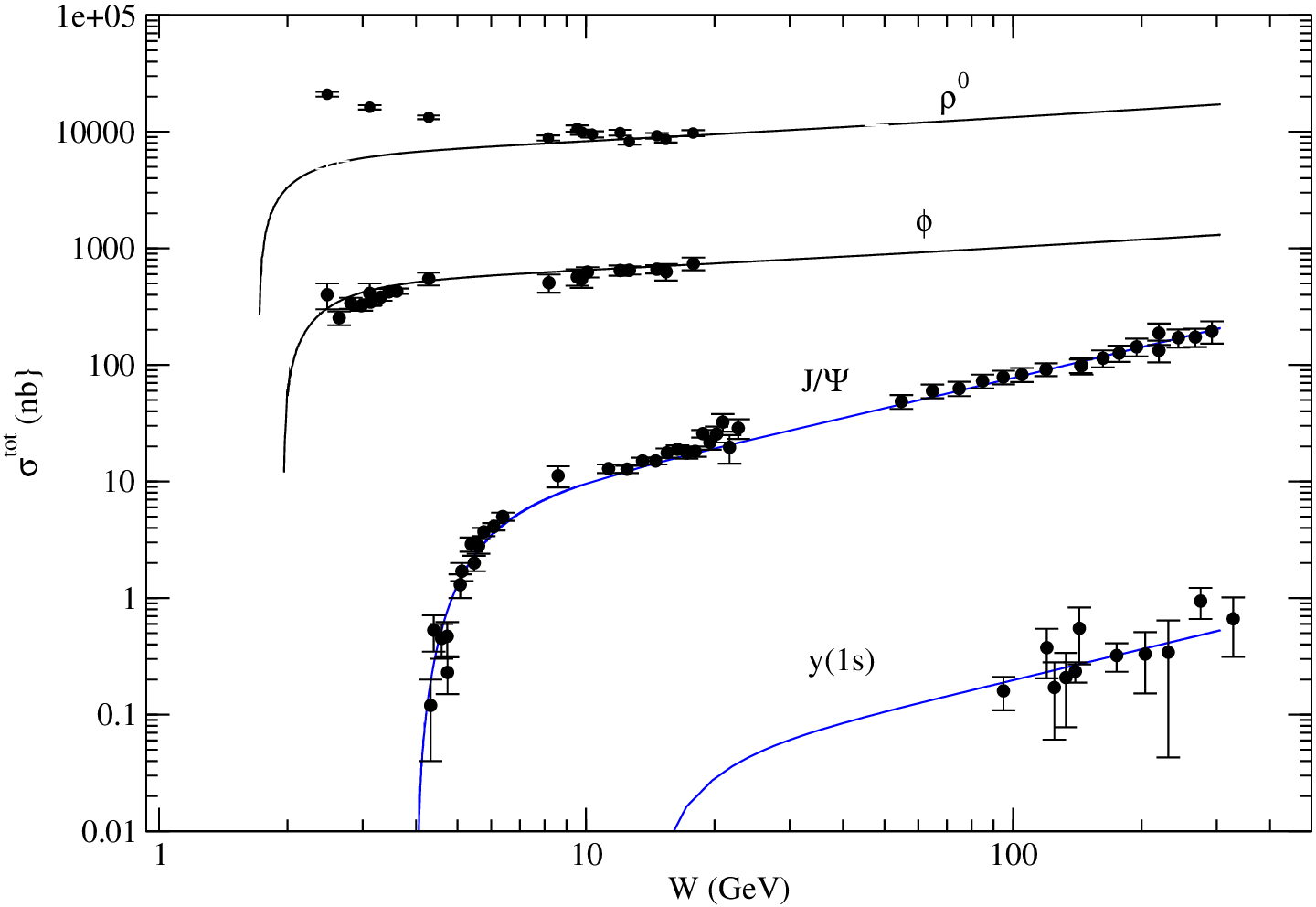}
\caption{Fits to the data of the total cross sections ($\sigma^{tot}$)
of photo-production of $\rho^0$, $\phi$, $J/\Psi$ and $y(1s)$ on the proton target.
Data are from Refs.\cite{jpsi-1}-\cite{phi-4}.}
\label{fig:totcrst-all-v}
\end{center}
\end{figure}

In Fig.\ref{fig:totcrst-all-v}, we  
 see that the data for the  $\phi$, $J/\Psi$, and $y(1s)$ production
can be described  very well by the Pomeron-exchange model. 
On the other hand, the $\rho$ production data at low energies clearly need
other mechanisms such as  the 
meson-exchange mechanisms illustrated in Ref.\cite{pich}.

 It appears that the slop parameter $\alpha_0$
for the energy-dependence of
the diffractive production of heavy quarks (
$c$ and $b$) is rather different from that for light quarks ($u$, $d$, $s$). It will be interesting to understand
this observation.
 
\section{VMD-LQCD model}

We now use the vector meson dominance (VMD) assumption
and  the $J/\Psi$-N potential  $v_{LQCD}$ of Ref.\cite{sasaki}  to construct
a model (VMD-LQCD) 
to predict the $J/\Psi$ photo-production 
cross sections.
It is defined by the following  Hamiltonian (from now on, we also use $V$ to denote $J/\Psi$):
\begin{eqnarray}
H = H_0 + v_{LQCD}(r) + \frac{em^2_V}{f_V}\int dx A_\mu(x)\phi_V^\mu(x)\,\,,
\label{eq:lqcd-h}
\end{eqnarray}
where $H_0$ is the free Hamiltonian, $f_{V=J/\Psi} =11.2$ as in the Pomeron-exchange model of section II, 
$A_\mu(x)$ and $\phi_{V}^\mu(x)$ are the field operators of the photon and the  considered vector meson, respectively.
Within the Hamiltonian formulation of hadron  reactions\cite{gw,feshbach,msl},
 the  amplitude
of $\gamma(\vec{q})+N(-\vec{q})\rightarrow J/\Psi(\vec{q})+N(-\vec{k})$ 
can then  be  written as 
\begin{eqnarray}
<\vec{k}\lambda_Vm_s|T_{VN,\gamma N}(W)|\vec{q}\lambda_\gamma m_s'>& =&
<\vec{k}\lambda_Vm_s|T_{LQCD}(W)|\vec{q}\lambda_\gamma m_s'>
\label{eq:lqcd-t}
\end{eqnarray}
where
\begin{eqnarray}
<\vec{k}\lambda_Vm_s|T_{LQCD}(W)|\vec{q}\lambda_\gamma m_s'>
&=&
 <\vec{k},\lambda_Vm_s|t_{VN,VN}(W)|\vec{q},\lambda_\gamma m'_s> 
\frac{1}{W-E_N(q)-E_V(k)+i\epsilon} \nonumber \\
&&\times
[\frac{em^2_V}{f_V}\frac{1}{(2\pi)^{3/2}}\frac{1}{\sqrt{2q}}\frac{1}{\sqrt{2E_V(q)}}]\,,
\label{eq:lqcd-t-1}
\end{eqnarray}
where  the outgoing vector meson  momentum $k=|\vec{k}|$ and the incoming photon momentum
$q=|\vec{q}|$  are defined by
$W=E_V(k)+E_N(k)=E_N(q)+q$.

The $V+N\rightarrow V+N$  scattering amplitude 
$<\vec{k},m_vm_s|t_{VN,V N}(E)|\vec{q},\lambda m'_s>$ in Eq.(\ref{eq:lqcd-t}) is calculated
from the potential $v_{LQCD}(r)$ by solving the Lippmann-Schwinger equation
\begin{eqnarray}
t_{VN,V N}(E) = v_{LQCD} + v_{LQCD}\frac{1}{W-H_0+i\epsilon}t_{VN,V N}(E)\,.
\label{eq:lpeq}
\end{eqnarray}
Note that $|\vec{q}|\neq |\vec{k}|$  and hence $<\vec{k},\lambda_Vm_s|t_{VN,V N}(E)|\vec{q},\lambda_\gamma m'_s>$
in Eq.(\ref{eq:lqcd-t}) is a half-off-shell
t-matrix and can not be directly determined by  the elastic scattering,
$V(\vec{k})  + N(-\vec{k}) \rightarrow V(\vec{k}^{\,\,'})  + N(-\vec{k}^{\,\,'}) $, cross sections
defined by
\begin{eqnarray}
\frac{d\sigma_{VN,VN}}{d\Omega}(W)
&=&\frac{(2\pi)^4}{k^2}\rho^2_{V,N}(k)\frac{1}{6}\sum_{\lambda_V,m_s}\sum_{\lambda'_V,m'_s}
|<\vec{k},\lambda_V m_s|t_{VN,VN}(W)|\vec{k}^{\,\,'}\lambda'_V m'_s>|^2 \nonumber \\
&&
\label{eq:crst-vnvn}
\end{eqnarray}
where $|\vec{k}^{\,\,'}|=|\vec{k}|=k$. 

In this work, we use $v_{LQCD}(r)$  extracted from a  LQCD calculation of  Ref.\cite{sasaki}.
Their LQCD data can be approximately  fitted\cite{sasaki-1} by 
\begin{eqnarray}
v_{LQCD}(r)= v_0 \, \frac{e^{-\alpha\,r}}{r}
\label{eq:v-lqcd}
\end{eqnarray}
We consider the ranges of parameters :
 $v_0= (-0.06, -0.11)$ and $ \alpha =( 0.3,  0.5)$ GeV, as estimated in Ref.\cite{sasaki-1}.
In left side of Fig.\ref{fig:totcrst-sasaki-lqcd}, we see that the LQCD data presented in Ref.\cite{sasaki}
 can be fitted
very well  with $v_0=-0.06$ and $\alpha= 0.3$ GeV ( pot-1)). However, the short-range part
at $r < $ about 0.4 fm is difficult\cite{sasaki-1}  to quantify in this LQCD calculation 
with a lattice spacing $\sim 0.1$ fm.
 Thus the potential (pot-2) with $v_0=-0.11$ and $\alpha= 0.5$ GeV  which fits
 only the data at $r >$ about 0.4 fm will also be considered in our calculations.

\begin{figure}[h] \vspace{+1.cm}
\begin{center}
\includegraphics[clip,width=0.4\textwidth,angle=0]{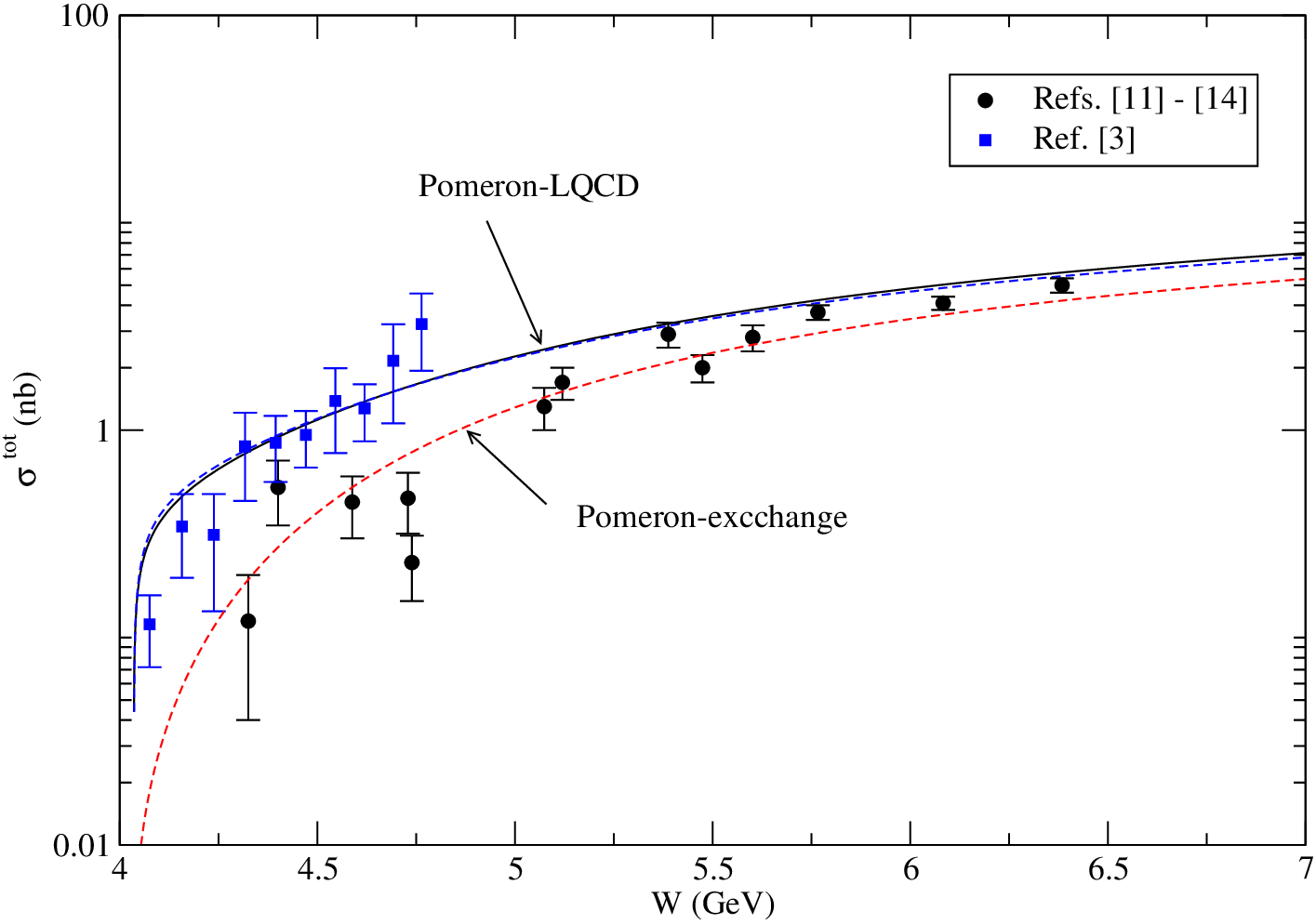}
\includegraphics[clip,width=0.4\textwidth,angle=0]{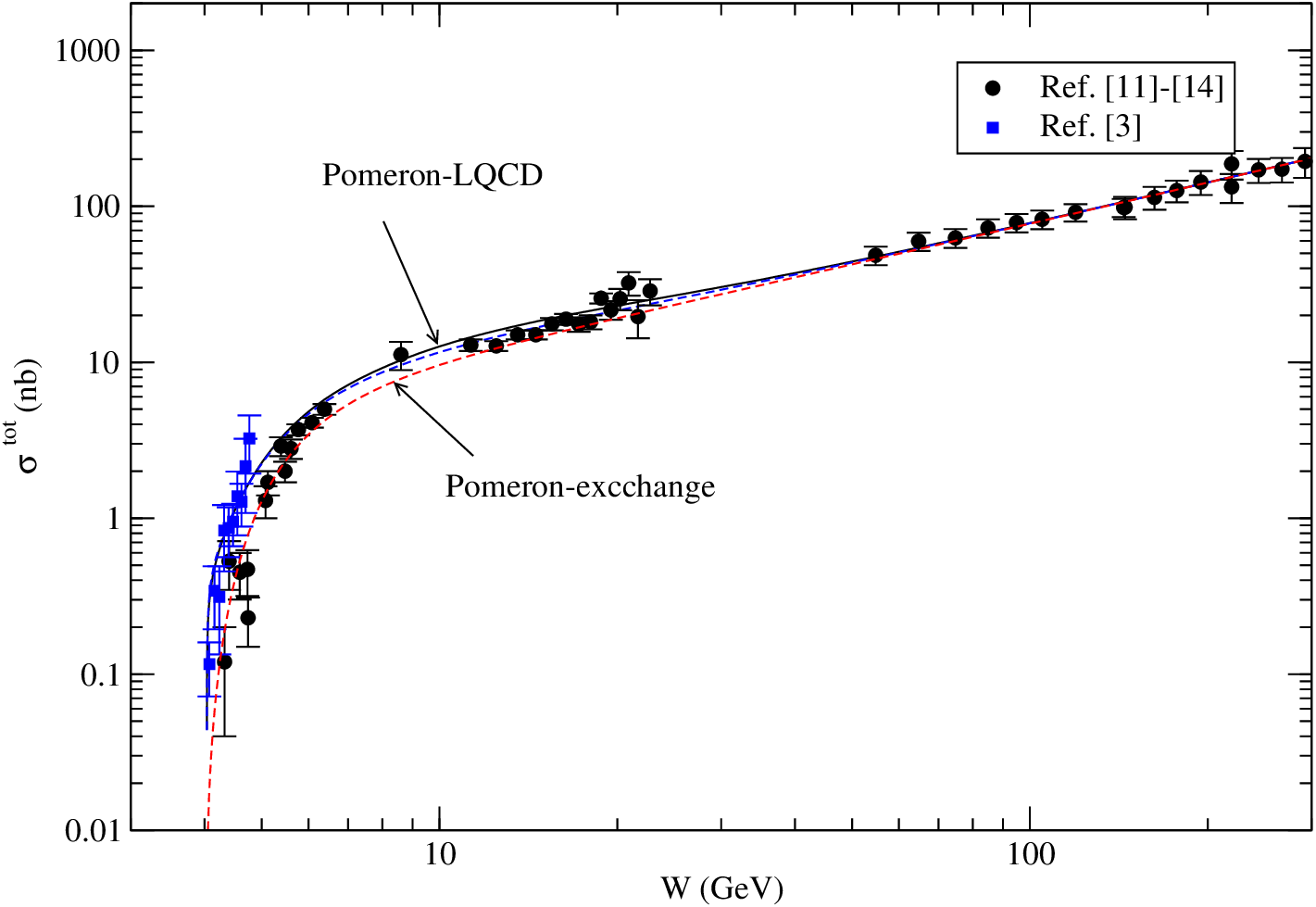}
\caption{Comparison of the  total cross sections calculated from Pomeron-exchange (dotted curves) and
Pomeron-LQCD model with $J/\Psi$-N potentials pot-1 (solid curves) and pot-2 (dashed curves).
The results from pot-1 (pot-2)  are obtained by multiplying  $F^{off}=0.75$ (0.41) to VMD coupling constant $1/f_V$
to fit the data and are alomst indistinguishable.
The data are from Refs.\cite{jpsi-1}-\cite{jpsi-4} and \cite{jlab}.}
\label{fig:totcrst-lqcd-pot-woff}
\end{center}
\end{figure}

By using Eq.(\ref{eq:v-lqcd}) to solve scattering equation Eq.({\ref{eq:lpeq}), we can get
the matrix elements of $t_{VN,VN}(W)$ for
evaluating $\gamma +N \rightarrow J/\Psi +N $ amplitude Eq.(\ref{eq:lqcd-t}) and
the differential cross sections Eq.(\ref{eq:crst-gnjn}).
The predicted $J/\Psi$ photo-production cross sections are compared with the  data in
the right side of Fig.\ref{fig:totcrst-sasaki-lqcd}. We see that the results from pot-1  are comparable to 
the JLab data, and  are higher than the results from the
Pomeron-exchange model (red dotted curve) in the near threshold region. 
The results (dashed curve)  from pot-2 are much larger than the data. 
This indicates  the importance of LQCD data in the  $ r< $ about 0.4 fm region . 
We will discuss this in section V.

In Fig.\ref{fig:totcrst-lqcd-pot-log}, we see that the total cross sections calculated from the VMD-LQCD model
using $J/\Psi$-N potentials pot-1  and pot-2 
are well  below the data  in the high energy region. 
Clearly, it is necessary to extend the VMD-LQCD model to include  the mechanisms of Pomeron-exchange model.
  
\begin{figure}[h] \vspace{+2.cm}
\begin{center}
\includegraphics[clip,width=0.4\textwidth,angle=0]{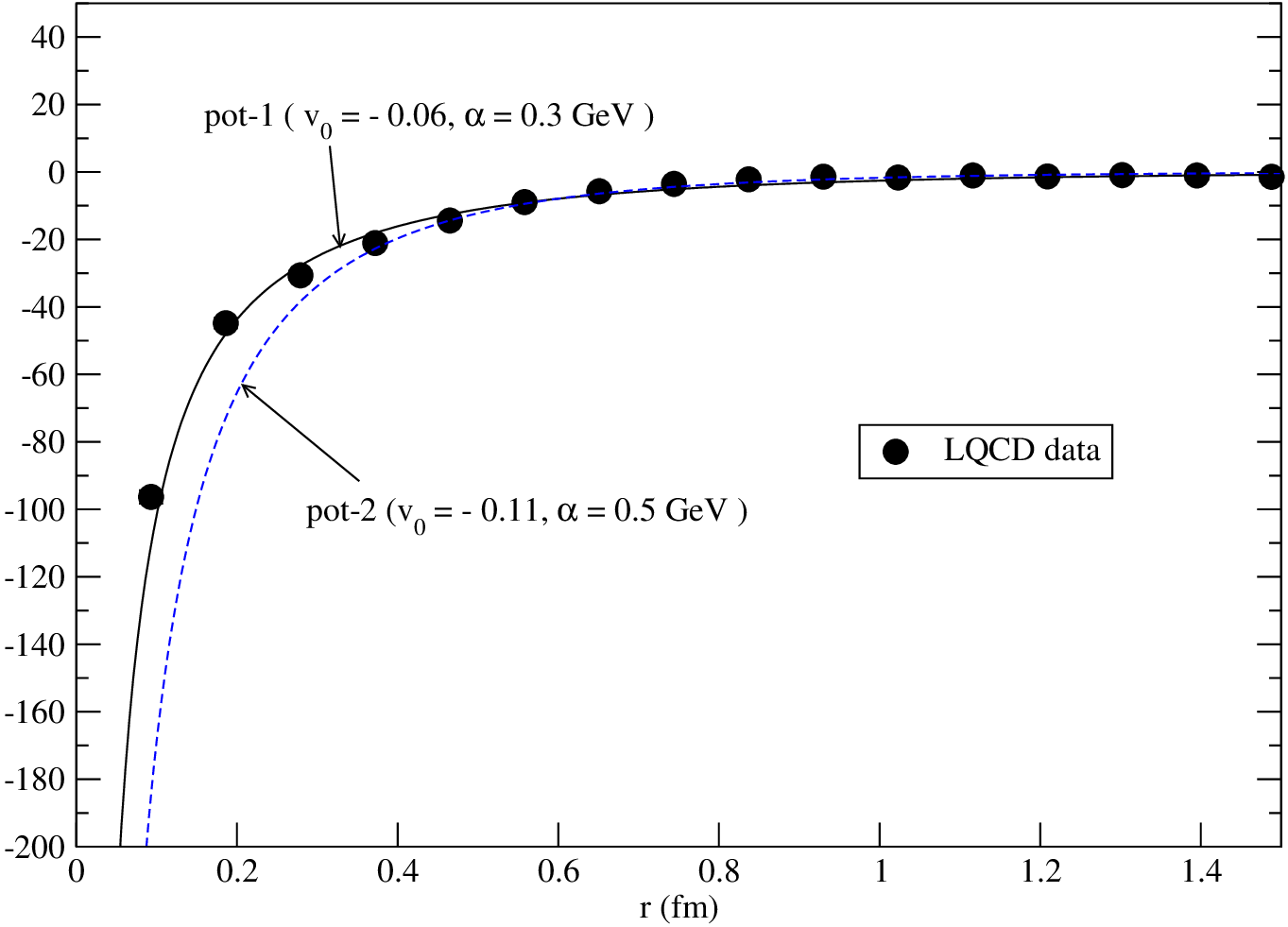}
\includegraphics[clip,width=0.4\textwidth,angle=0]{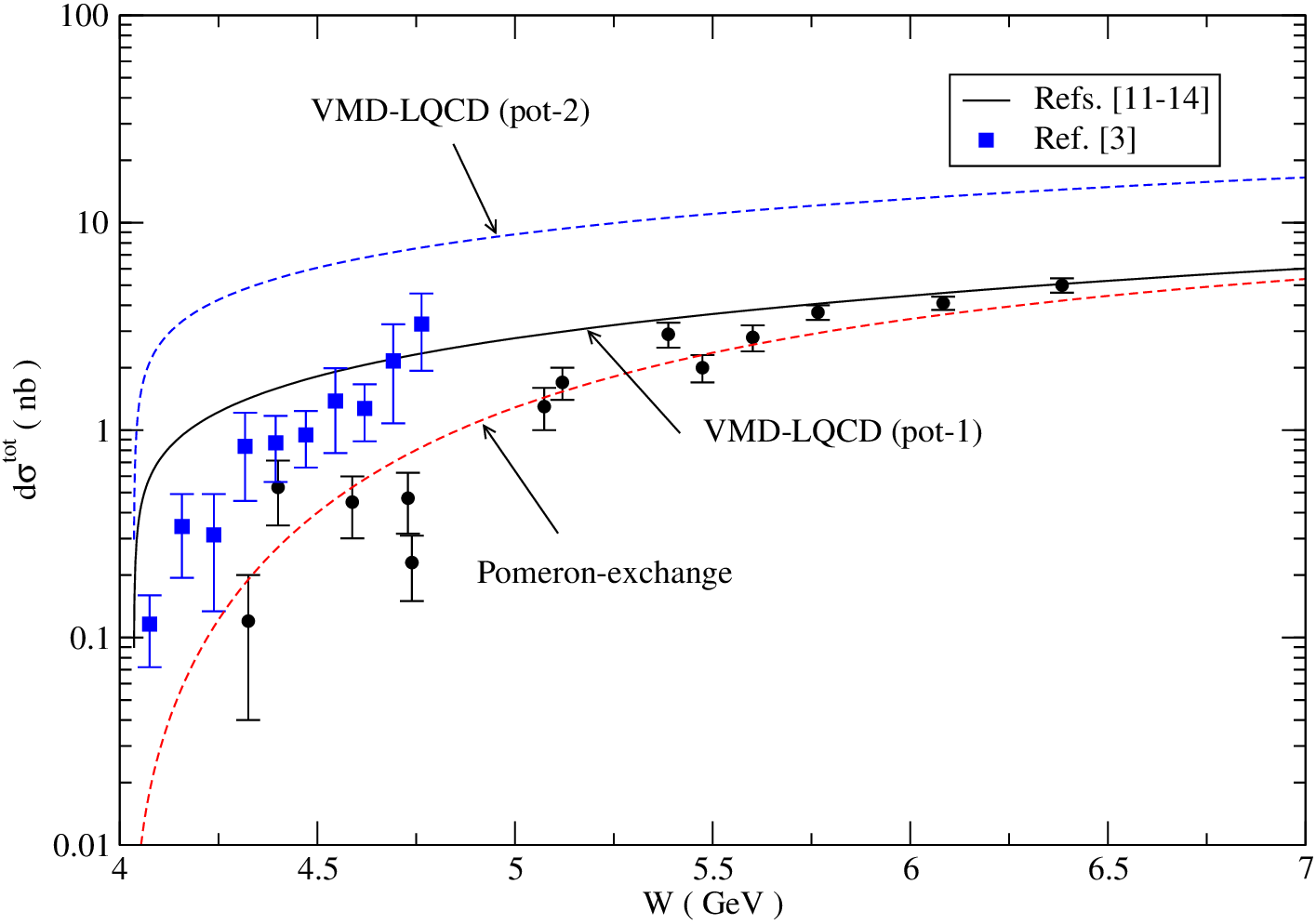}
\caption{Left: fits to the LQCD data of $J/\Psi$-N potential  of Ref.\cite{sasaki,sasaki-1}.
$v_0$, and $\alpha$ are the parameters of the potential Eq.(\ref{eq:v-lqcd}).
 Right: the 
total cross sections of  $\gamma +p \rightarrow J/\Psi + p$  calculated from VMD-LQCD models with 
$J/\Psi$-N potentials pot-1 (solid curve) and pot-2 (dashed curve)   are compared
with the data and the results (dotted curve) from the Pomeron-exchange model presented in section II.
The data are from Refs.\cite{jpsi-1}-\cite{jpsi-4} and \cite{jlab}.}
\label{fig:totcrst-sasaki-lqcd}
\end{center}
\end{figure}

\begin{figure}[h] \vspace{+2.cm}
\begin{center}
\includegraphics[clip,width=0.6\textwidth,angle=0]{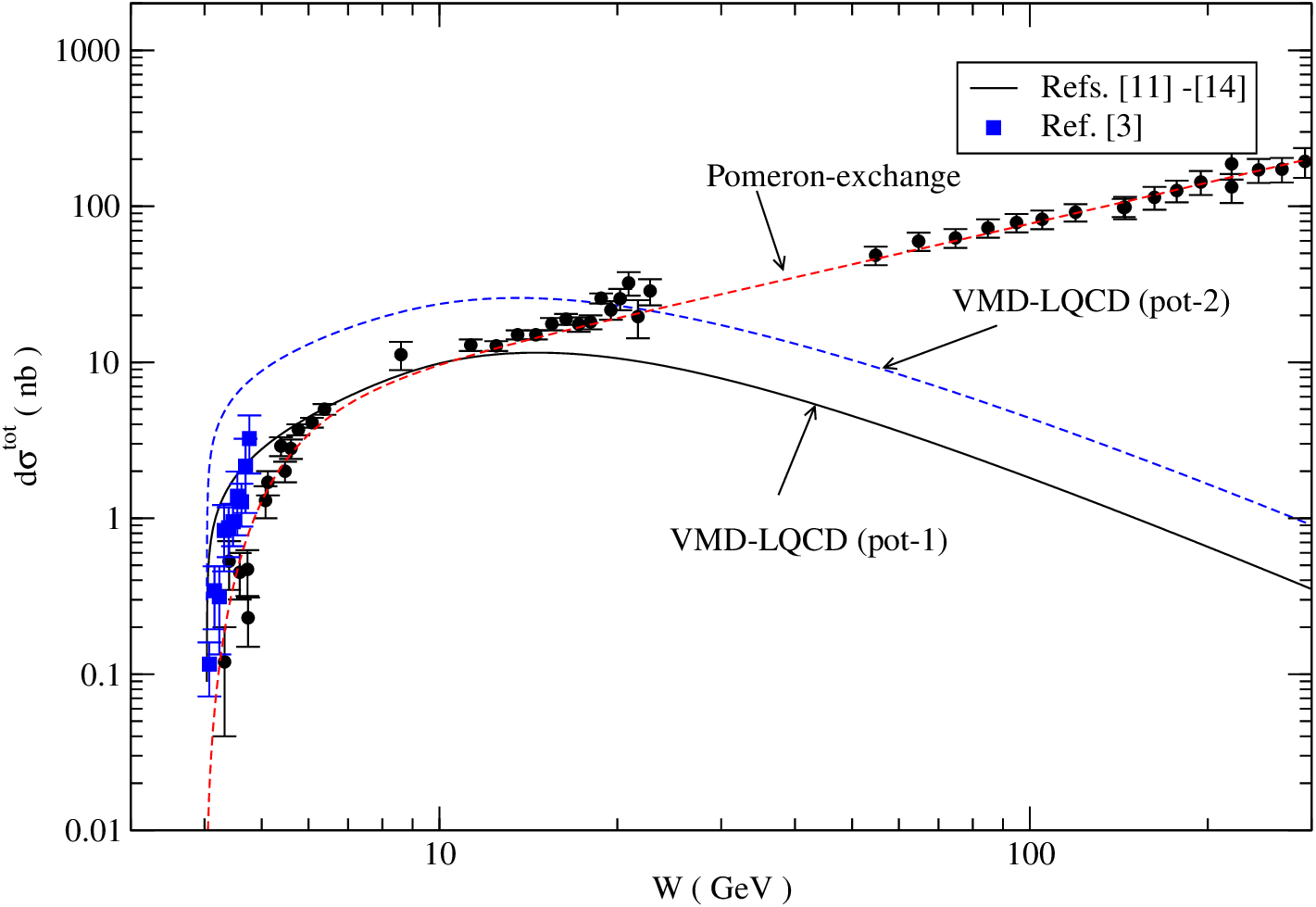}
\caption{Same as the right-side of Fig.\ref{fig:totcrst-sasaki-lqcd}, except also including
 the comparisons  with the data at high energies.}
\label{fig:totcrst-lqcd-pot-log}
\end{center}
\end{figure}

\section{Pomeron-LQCD model}

The  Pomeron-exchange model developed in Refs.\cite{ohlee,wulee} and used 
here is based on a "perturbative" analysis of
 Donnachie and Landshoff\cite{dl}. 
Its mechanism is therefore very  different from  $J/\Psi$-N potential extracted from
a LQCD calculation which account for the 
"non-perturbative" gluonic interactions between $J/\Psi$ and nucleon.
Following the well-established  approach in developing models of hadron-hadron scattering,
we now extend the Hamiltonian Eq.(\ref{eq:lqcd-h}) to 
develop a model which contain  two  mechanisms:(1) the "non-perturbative" $v_{LQCD}$ extracted 
from LQCD calculation, 
 (2) the "perturbative"  two-gluon-exchange amplitudes of Donnachie and Landshoff which can be generated
 by a potential $v_{PQCD}$.
The $J/\Psi$ photo-production is then defined by the following Hamiltonian:
\begin{eqnarray}
H = H_0 + [\,v_{LQCD}(r)+v_{PQCD}\,] + \frac{em^2_V}{f_V}\int dx A_\mu(x)\phi_V^\mu(x)
\label{eq:lqcd-h1}
\end{eqnarray}
By using the two-potential
formula of the well-established reaction theory\cite{gw}, the amplitude of $J/\Psi$ photo-production
derived from Eq.(\ref{eq:lqcd-h1}) 
is of the same form of Eq.(\ref{eq:lqcd-t}) except that the $V+N \rightarrow V+N$ amplitude
is replaced by
\begin{eqnarray}
<\vec{k}|t_{VN,V N}(W)|\vec{q}>\,\, \rightarrow\,\,
 <\vec{k}|t_{VN,V N}(W)|\vec{q}> +<\vec{k}|t^{PQCD}_{VN,VN}(W)|\vec{q}>\,
\label{eq:pom-lqcd-vnvn}
\end{eqnarray}
where $t_{VN,V N}(W)$ is defined by Eq.(\ref{eq:lpeq}), and 
\begin{eqnarray}
<\vec{k}|t^{PQCD}_{VN,VN}(W)|\vec{q}> = <\phi^{(-)}_{\vec{k},W}|v_{PQCD}|\Psi^{(+)}_{\vec{q},W}>\,,
\label{eq:pom-lqcd-t-1}
\end{eqnarray}
with
\begin{eqnarray}
<\phi^{(-)}_{\vec{k},W}|&=&<\vec{k}|[1+ v_{LQCD}\frac{1}{W-H_0 - v_{LQCD}+i\epsilon}] \nonumber \\
&=&<\vec{k}|[1+ t_{VN,VN}(W)\frac{1}{W-H_0 + i\epsilon}] 
\label{eq:phim}
\end{eqnarray}
and
\begin{eqnarray}
|\Psi^{(+)}_{\vec{q},W}> = [1 + \frac{1}{W - H_0-v_{LQCD}-v_{PQCD} +i\epsilon}]
(v_{LQCD}+v_{PQCD})|\vec{k}>\,.
\label{eq:phip}
\end{eqnarray}
With the replacement Eq.(\ref{eq:pom-lqcd-vnvn}), the photo-production amplitude can then be
written as
\begin{eqnarray}
<\vec{k}\lambda_Vm_s|T_{VN,\gamma N}(W)|\vec{q}\lambda_\gamma m_s'>& =&
<\vec{k}\lambda_Vm_s|T_{LQCD}(W)|\vec{q}\lambda_\gamma m_s'> \nonumber \\
&& +
<\vec{k}\lambda_Vm_s|T_{PQCD}(W)|\vec{q}\lambda_\gamma m_s'>
\label{eq:pom-lqcd-f}
\end{eqnarray}
where $<\vec{k}\lambda_Vm_s|T_{LQCD}(W)|\vec{q}\lambda_\gamma m_s'>$ is defined by Eq.(\ref{eq:lqcd-t-1}), and
\begin{eqnarray}
<\vec{k}\lambda_Vm_s|T_{PQCD}(W)|\vec{q}\lambda_\gamma m_s'>
&=&
 <\vec{k},\lambda_Vm_s|t^{PQCD}_{VN,VN}(W)|\vec{q},\lambda_\gamma m'_s>
\frac{1}{W-E_N(q)-E_V(k)+i\epsilon} \nonumber \\
&&\times
[\frac{em^2_V}{f_V}\frac{1}{(2\pi)^{3/2}}\frac{1}{\sqrt{2q}}\frac{1}{\sqrt{2E_V(q)}}]\,,
\label{eq:lqcd-t-2}
\end{eqnarray}

In the absence of a model of $v_{PQCD}$ for solving Eqs.(\ref{eq:pom-lqcd-t-1})-(\ref{eq:phip}),
we assume that the amplitude $<\vec{k}|T_{PQCD}(W)|\vec{q}>$  
can be identified with the amplitude of Eq.(\ref{eq:pomt}) of the 
the Pomeron-exchange model described in section II:
\begin{eqnarray}
<\vec{k}\lambda_Vm_s|T_{PQCD}(W)|\vec{q}\lambda_\gamma m_s'>\,\rightarrow\,
<\vec{k}\lambda_Vm_s|T_P(W)|\vec{q}\lambda_\gamma m_s'>.
\label{eq:pqcd-app}
\end{eqnarray}

The  model
 defined by Eqs.(\ref{eq:lqcd-h1})- Eq.(\ref{eq:pqcd-app})
will be refereed to as Pomeron-LQCD model within which
 the photo-production amplitude is 
then calculated  by using the following form
\begin{eqnarray} 
<\vec{k}\lambda_Vm_s|T_{VN,\gamma N}(W)|\vec{q}\lambda_\gamma m_s'>& =&
<\vec{k}\lambda_Vm_s|T_{LQCD}(W)|\vec{q}\lambda_\gamma m_s'> \nonumber \\
&&+<\vec{k}\lambda_Vm_s|T_P(W)|\vec{q}\lambda_\gamma m_s'>
\label{eq:pom-lqcd-f}
\end{eqnarray}
where $<\vec{k}\lambda_Vm_s|T_{LQCD}(W)|\vec{q}\lambda_\gamma m_s'>$ and
$<\vec{k}\lambda_Vm_s|T_P(W)|\vec{q}\lambda_\gamma m_s'>$ can be calculated by using Eq.(\ref{eq:lqcd-t-1})
and Eq.(\ref{eq:pomt}), respectively.

By using Eq.(\ref{eq:pom-lqcd-f}),  the $J/\Psi$ photo-production total cross sections are compared
with the data in Fig.\ref{fig:totcrst-lqcd-pot}. The results from keeping only $T_{LQCD}$ and $T_P$ are
also shown for comparisons.
In  the left-side of Fig.\ref{fig:totcrst-lqcd-pot},
we see that 
  the contrinution  from the amplitude $T_{LQCD}$
dominants the cross sections at low energies. However, it is significantly  larger than three JLab data in 
$W < 4.3$ GeV near threshold.
  At higher energies, it interferes coherently with the Pomeron-exchange contribution (red dashed curve)
to give cross sections a little higher than the old data.
If pot-2 of $v_{LQCD}$ shown in the left of
Fig.\ref{fig:totcrst-sasaki-lqcd} is used,  the calculated total cross sections
are a factor of about 5 larger than the JLab data,
 similar to that (blue dashed curve)
shown in the right side of  Fig.\ref{fig:totcrst-sasaki-lqcd}.

At high energies, the perturbative amplitude $T_P$ dominants and  the Pomeron-LQCD model can describe the
data as good as the Pomeron-exchange model described in section II. This is shown in the right-side of 
Fig.\ref{fig:totcrst-lqcd-pot}.

We now observe that 
the best agreements with both the JLab data and earlier data can be obtained by
multiplying  the VMD constant $1/f_V$ in the amplitude Eq.(\ref{eq:lqcd-t}})
by a factor $F^{off}=0.75$ (0.41) for the calculations using pot-1 (pot-2).
These fits are shown in
\ref{fig:totcrst-lqcd-pot-woff}.  For consistency, the VMD constant in the pomeron-exchange amplitude
Eq.(\ref{eq:pom-a}) should also be multiplied by the same $F^{off}$ factor within 
Pomeron-LQCD model. This however can be interpreted as just re-defining the Pomeron-quark coupling constant
$\beta_c \rightarrow \beta_c/F^{off}$. We will discuss this $F^{off}$ in the next section.

\begin{figure}[h] \vspace{+1.cm}
\begin{center}
\includegraphics[clip,width=0.4\textwidth,angle=0]{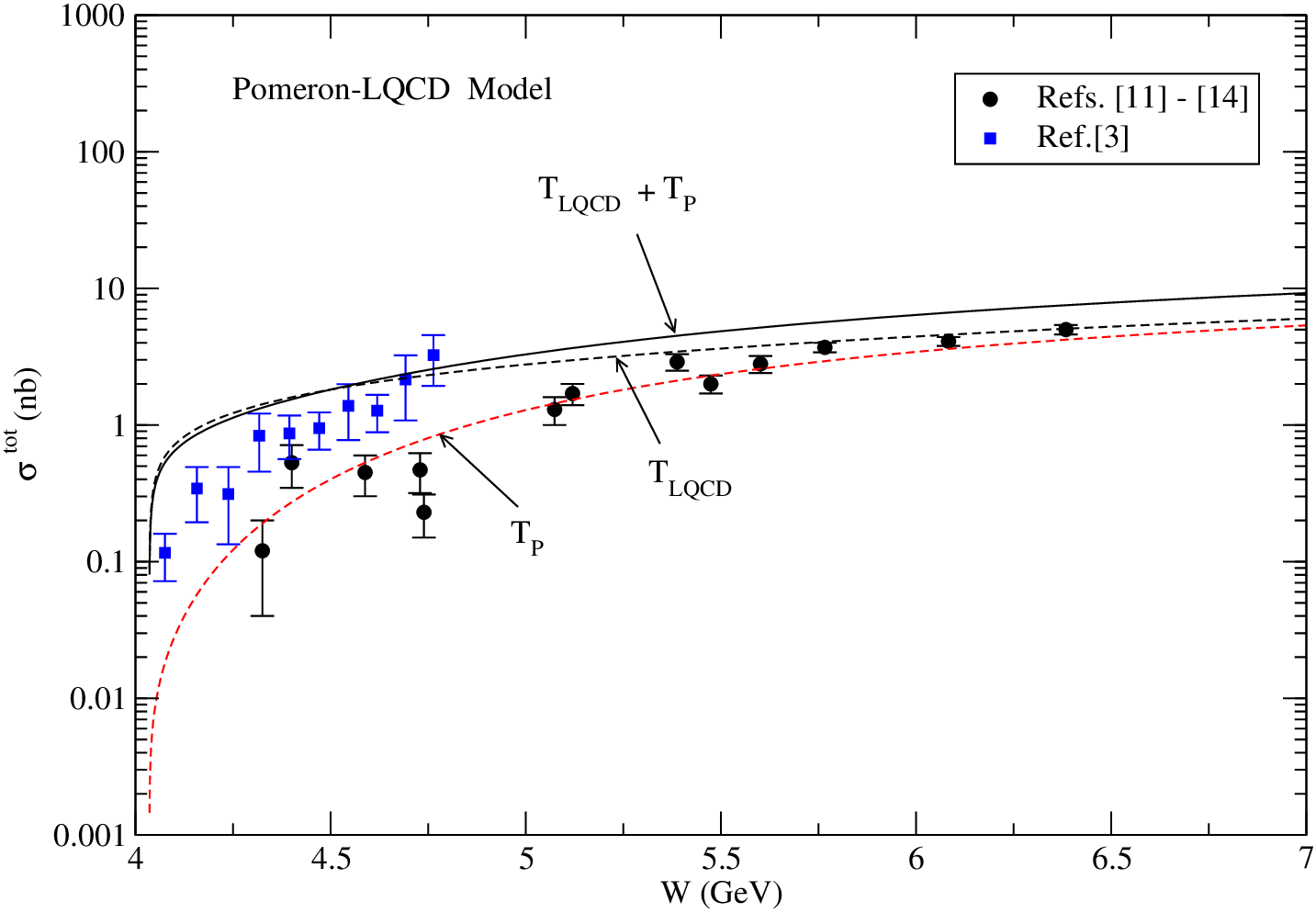}
\includegraphics[clip,width=0.4\textwidth,angle=0]{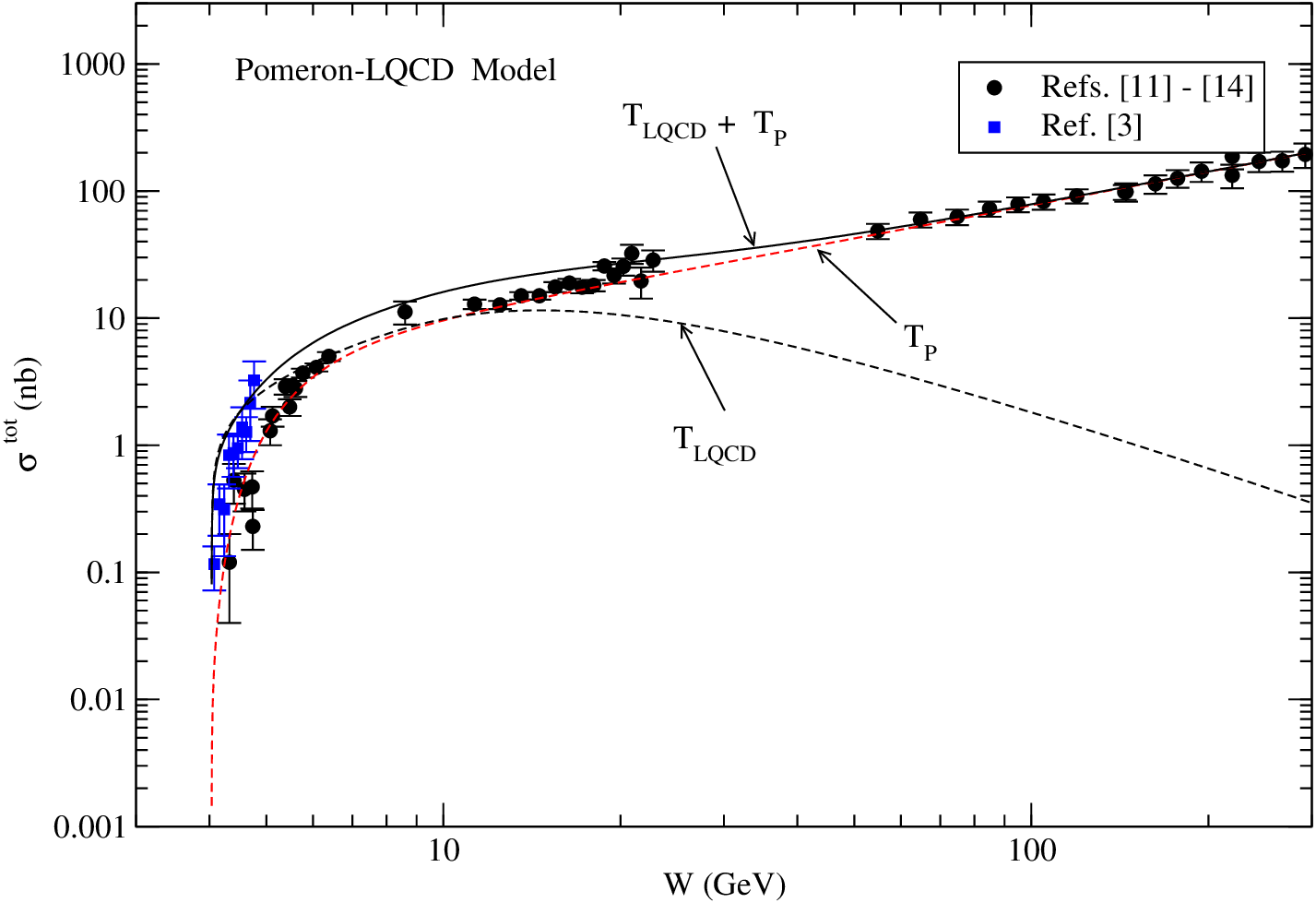}
\caption{The results (solid curves) from Pomeron-LQCD model are compared with the
data. The results from keeping only the amplitude $T_{LQCD}$ (blue dashed curves)and
$T_P$ (red dashed curves) of Eq.(\ref{eq:pom-lqcd-f}) are also shown.
 Left: from threshold to  $W=7$ GeV,
Right: from threshold to $W=300$ GeV. 
The data are from Refs.\cite{jpsi-1}-\cite{jpsi-4} and \cite{jlab}.}  
\label{fig:totcrst-lqcd-pot}
\end{center}
\end{figure}

\begin{figure}[h] \vspace{+2.cm}
\begin{center}
\includegraphics[clip,width=0.4\textwidth,angle=0]{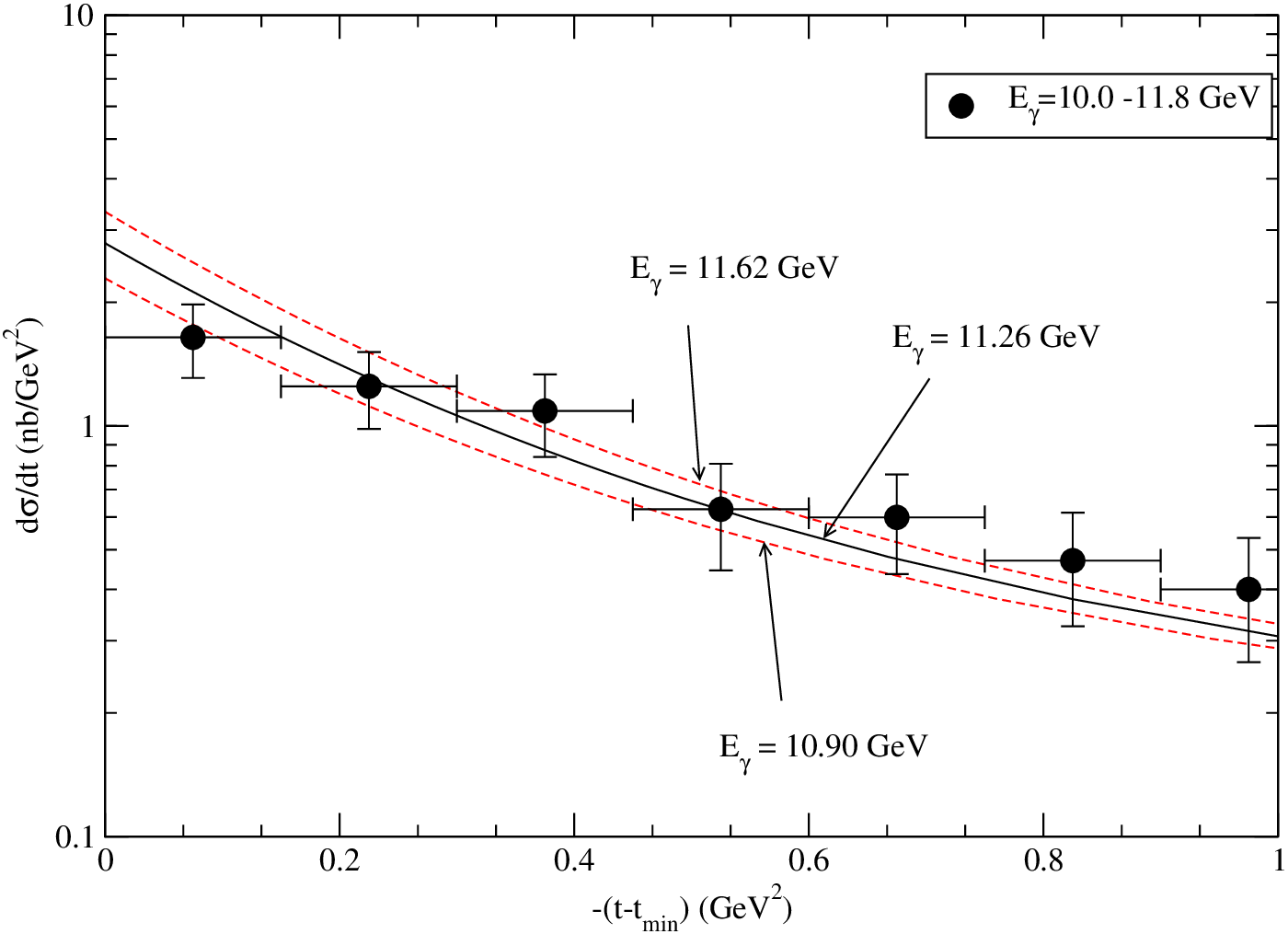}
\includegraphics[clip,width=0.4\textwidth,angle=0]{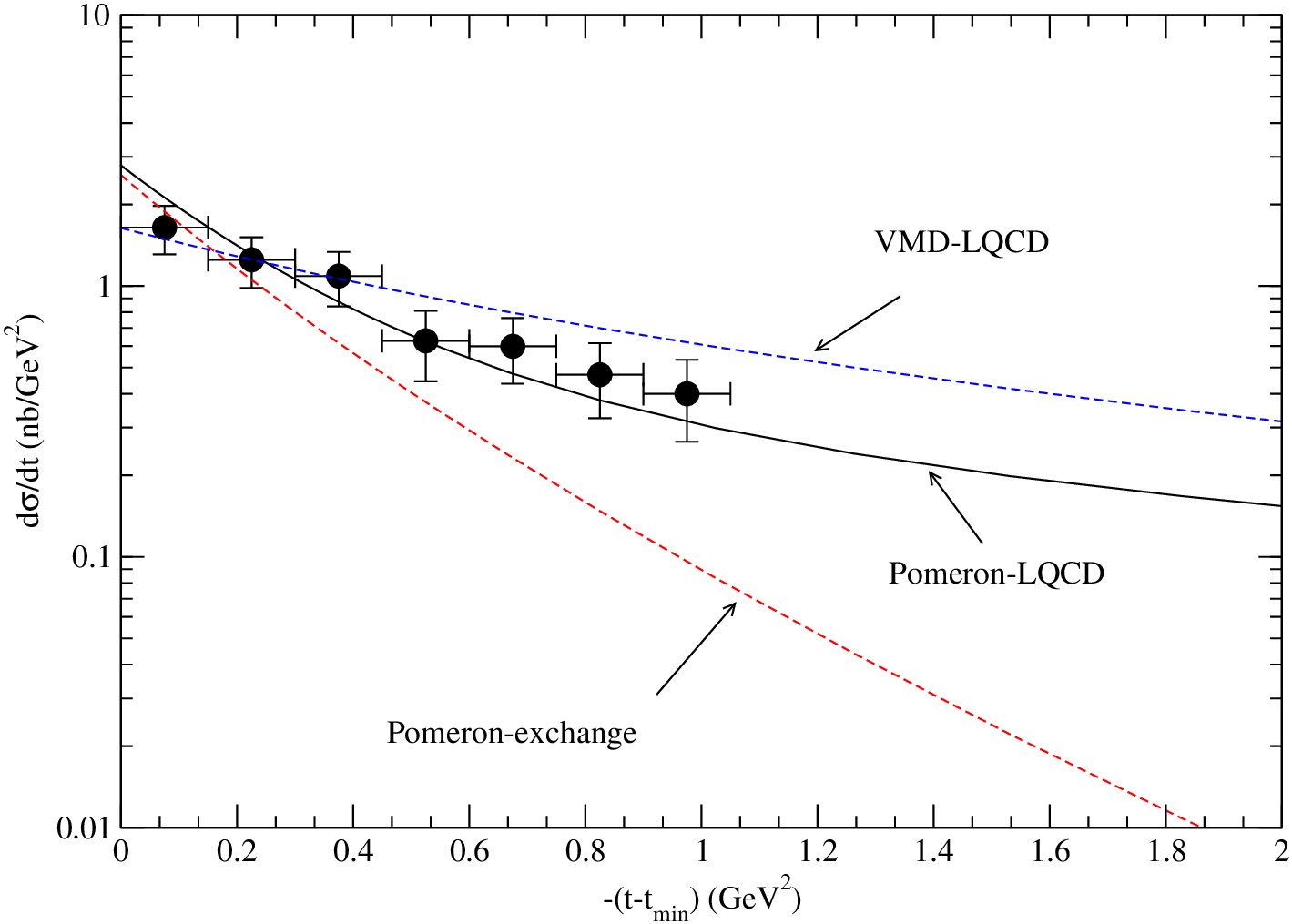}
\caption{Left: Differential cross sections of $\gamma + p \rightarrow J/\Psi+p$ 
 calculated from the  Pomeron-LQCD
model are compared with the JLab data\cite{jlab}. 
Right:Differential cross sections of $\gamma + p \rightarrow J/\Psi+p$  
 calculated from the  Pomeron-LQCD, VMD-LQCD, and Pomeron-exchange 
models are compared with the JLab data\cite{jlab}.}
\label{fig:dsdt-0}
\end{center}
\end{figure}

A better way to test the model is to compare the predicted differential cross section
$d\sigma/dt $ with the JLab data.
In the left  side of
Fig.\ref{fig:dsdt-0}, we see that the results from using $J/\Psi$-N
 potential pot-1 at three energies
in the range of JLab data agree well with the data.
In the right side of the same figure, we see that
Pomeron-LQCD model can describe the data much better than VMD-LQCD and Pomeron-exchange 
models. 

\begin{figure}[h] \vspace{+2.cm}
\begin{center}
\includegraphics[clip,width=0.4\textwidth,angle=0]{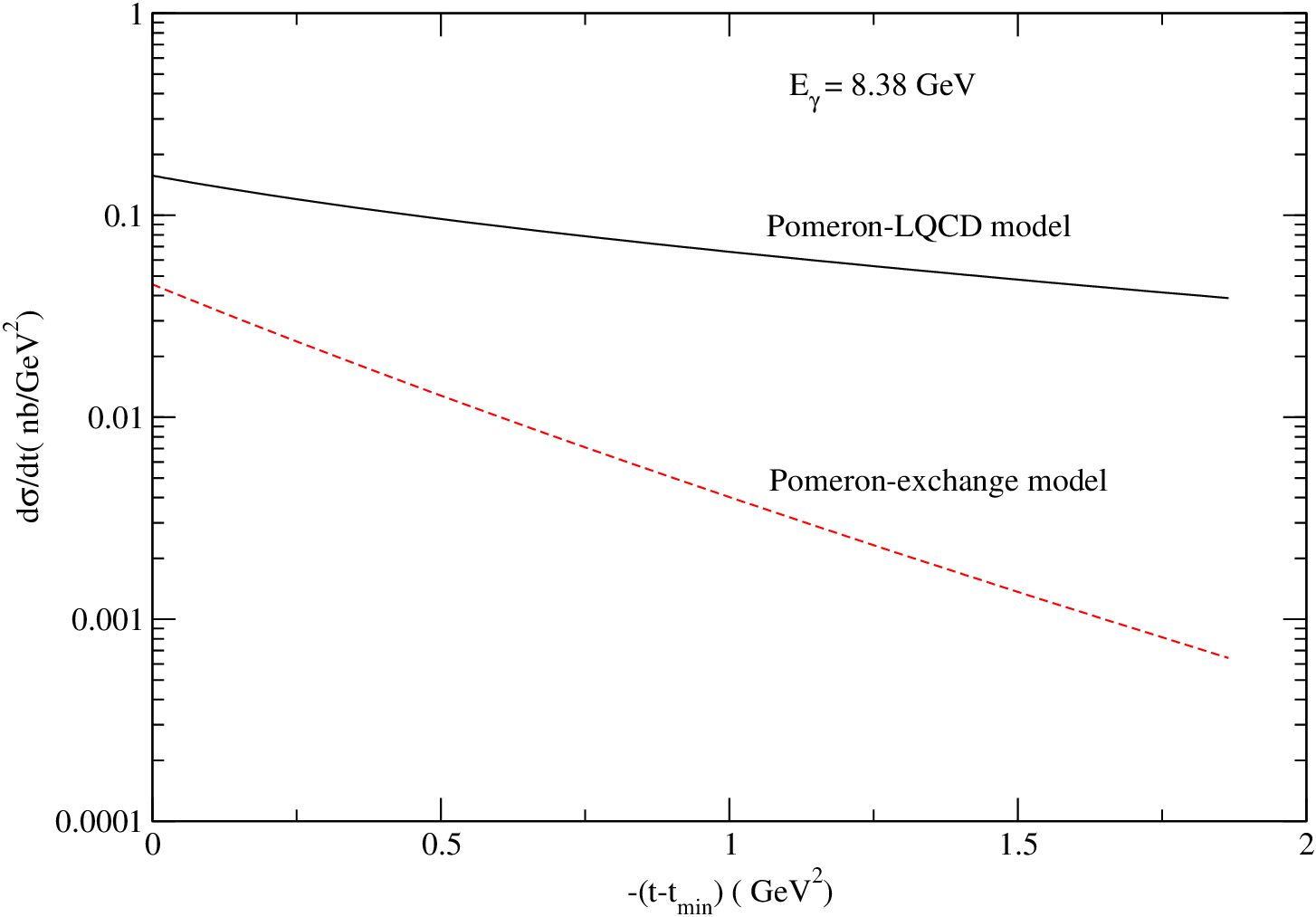}
\includegraphics[clip,width=0.4\textwidth,angle=0]{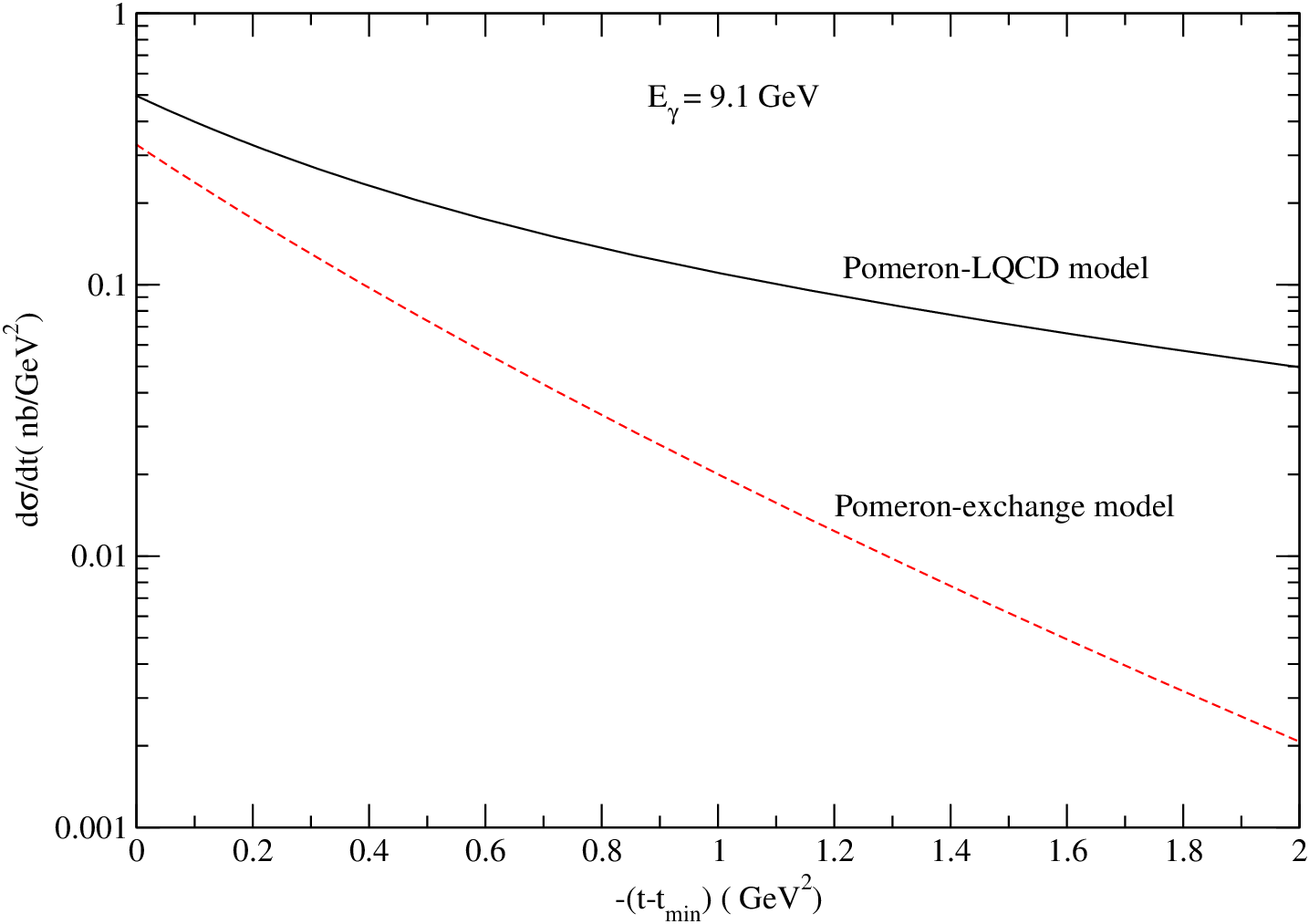}
\includegraphics[clip,width=0.4\textwidth,angle=0]{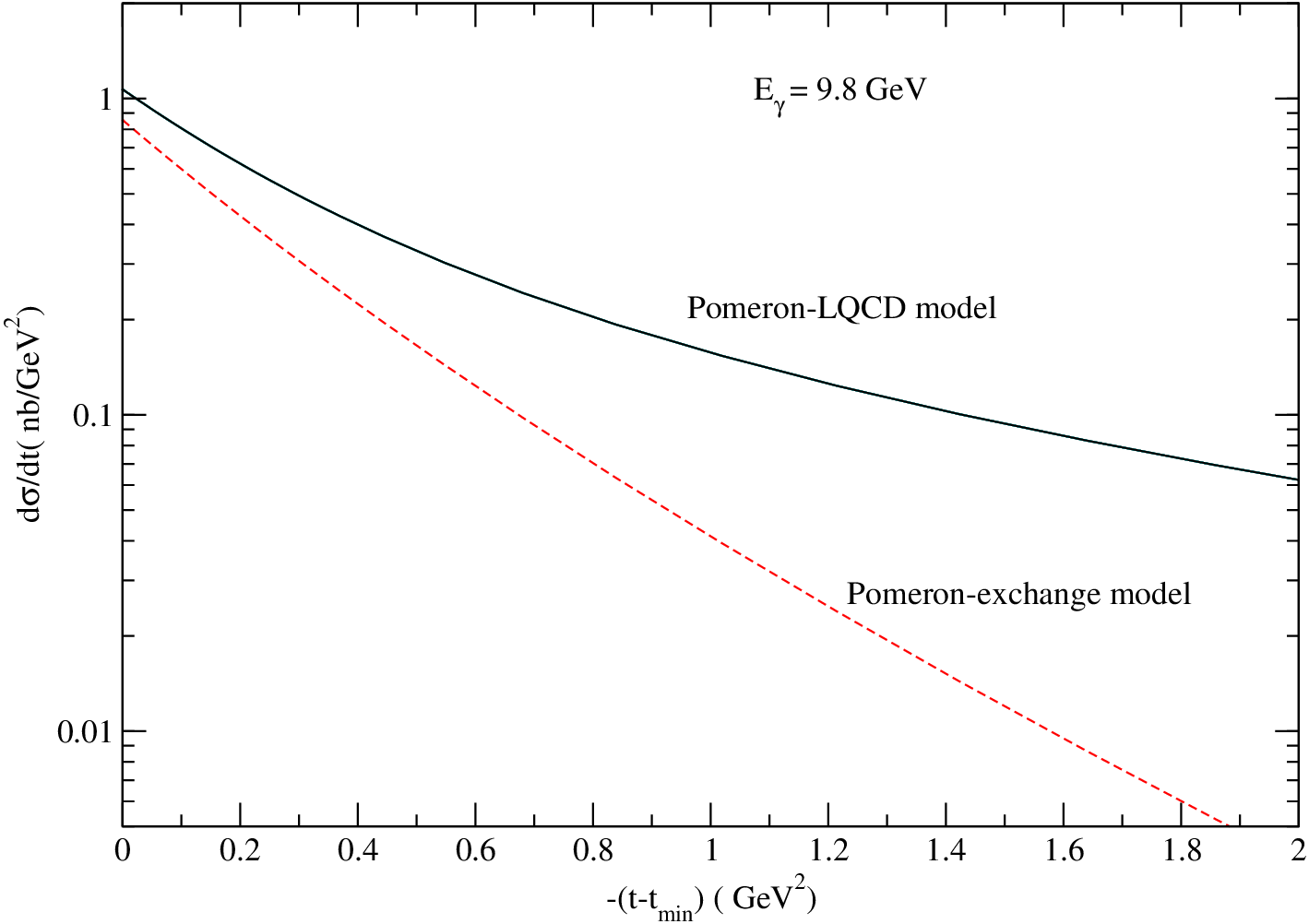}
\includegraphics[clip,width=0.4\textwidth,angle=0]{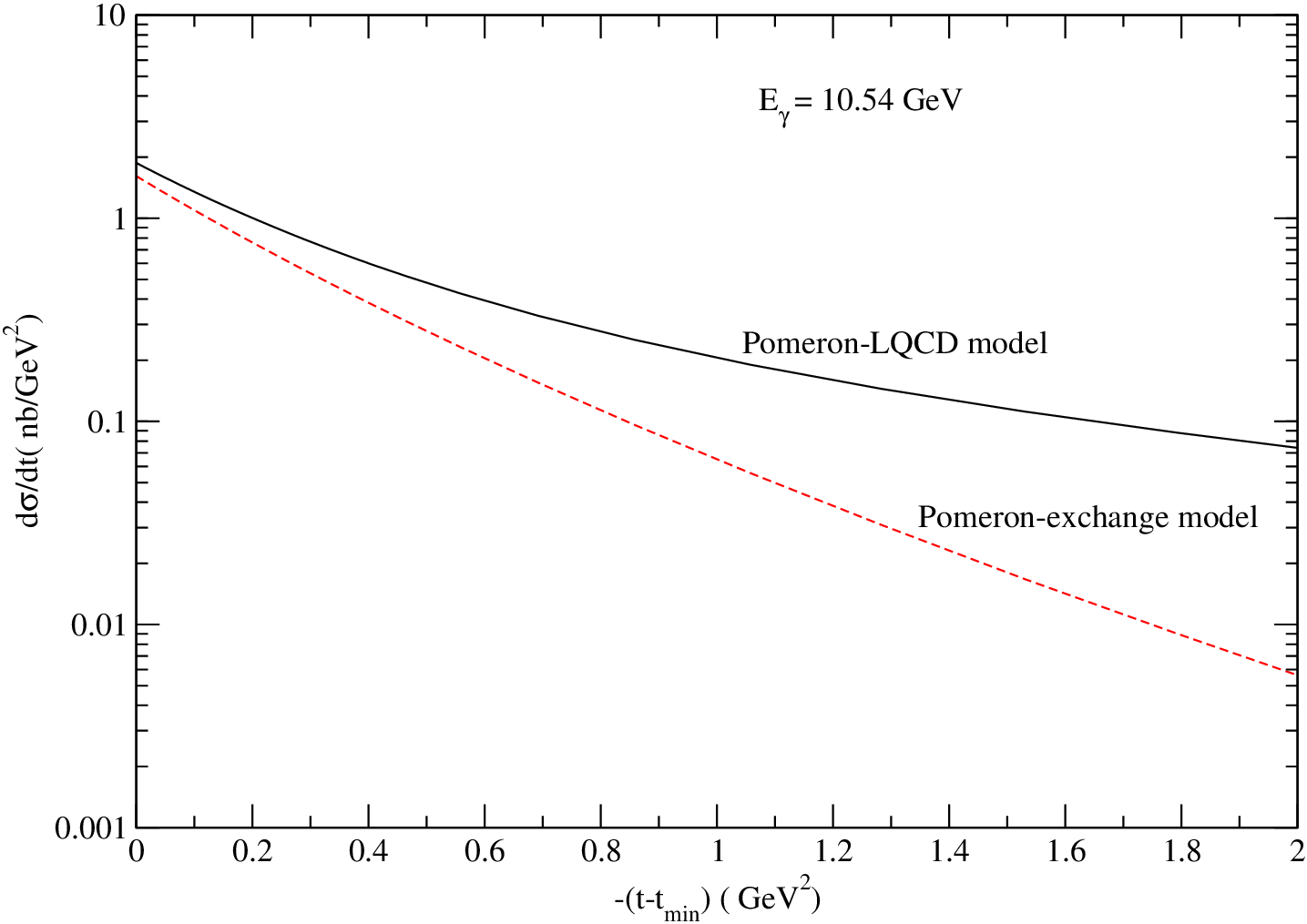}
\caption{Differential cross sections $\gamma + p \rightarrow J/\Psi +p$ calculated 
from the  Pomeron-LQCD model and Pomeron-exchange model are compared.
} 
\label{fig:dsdt-1}
\end{center}
\end{figure}

It will be interesting to test the predictions from Pomeron-LQCD model at  energies near threshold.
Our predictions at four energies of JLab data are shown in
Fig.\ref{fig:dsdt-1}.
We see that the predictions from Pomeron-exchange model (dashed curves) and Pomeron-LQCD model (solid curves)
have rather different $t$-dependence. Hopefully these differences can be tested by
the forthcoming data.

\section{Discussions and necessary improvements}

The good agreements with the data shown in Fig.\ref{fig:totcrst-lqcd-pot-woff}
 are obtained by multiplying the VMD coupling constant $1/f_V$ by
a factor $F^{off}=  075$ and 0.41 for the calculations using $J/\Psi$-N potential pot-1 and pot-2
shown in the left side of Fig.\ref{fig:totcrst-sasaki-lqcd}, respectively.
Here we note that the
 parameter $1/f_V$ in Eqs.(\ref{eq:pom-a}) for Pomeron-exchange model and
Eq.(\ref{eq:lqcd-h1}) for Pomeron-LQCD model   is conventionally determined
by $V\rightarrow e^+e^{-}$ decay width. Thus this coupling is for the  photon with $q^2=m^2_V$
which is different from
$q^2=0$ for the photo-production process considered in this work.
It is therefore reasonable to consider that
 $F^{off}$ is needed phenomenologically to account for this $q^2$-dependence  of VMD.
Since $m^2_{J/\Psi} \sim 9 $ GeV$^2$ is far away from $q^2=0$, the factor $F^{off}$ should deviate
significantly  from
1 and thus the results from pot-2 with a smaller $F^{off}=0.41$ 
 is more reasonable than those from pot-1.
If this speculation is correct, a $J/\Psi$-N potential which is
more attractive than the data presented in  Ref.\cite{sasaki}
is more consistent with Pomeron-LQCD model developed in this work. It will be interesting to have
a  LQCD calculation which  can reduce the uncertainties illustrated in 
Fig.\ref{fig:totcrst-sasaki-lqcd}. 

\begin{figure}[t] \vspace{+2.cm}
\begin{center}
\includegraphics[clip,width=0.4\textwidth,angle=0]{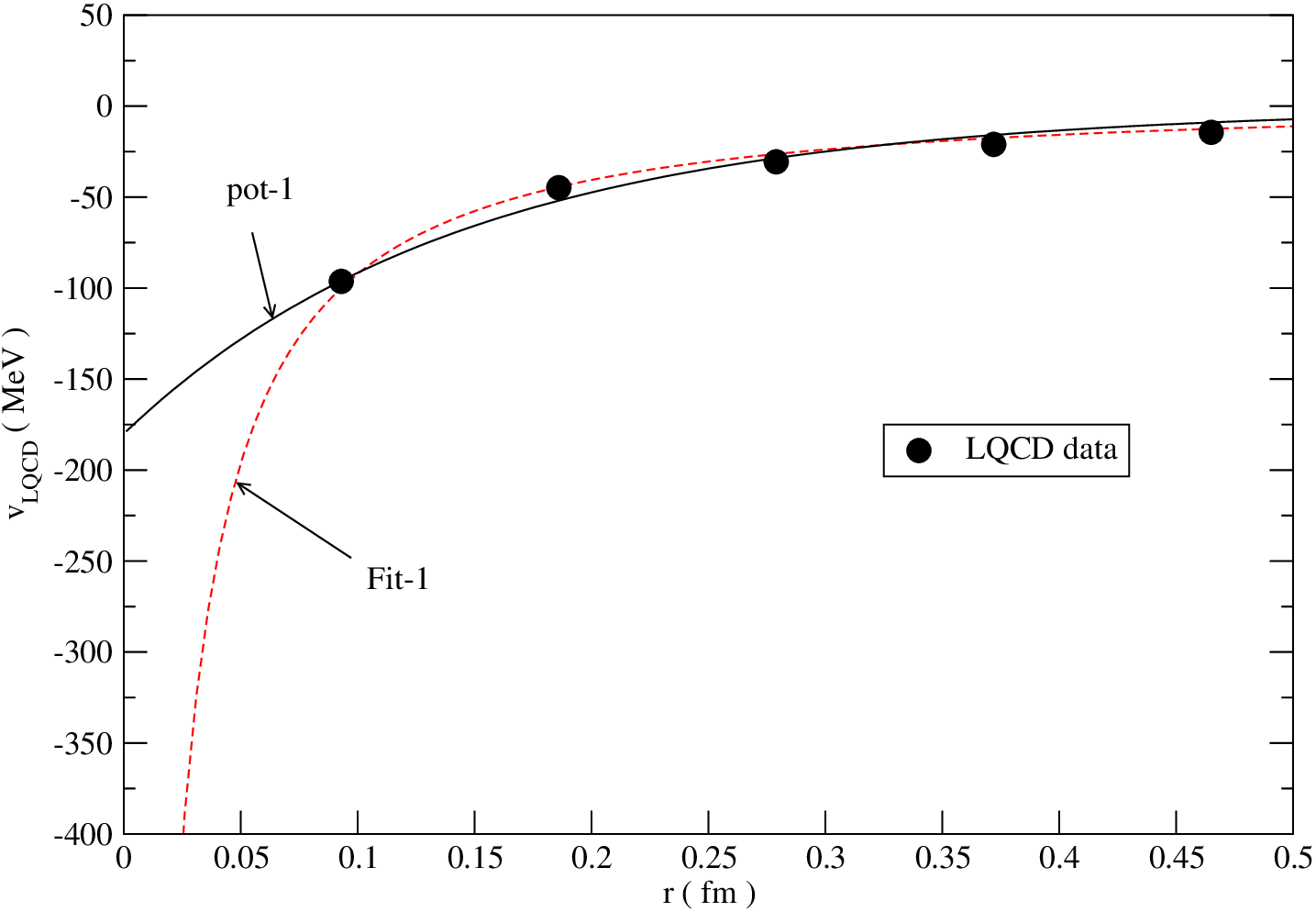}
\includegraphics[clip,width=0.4\textwidth,angle=0]{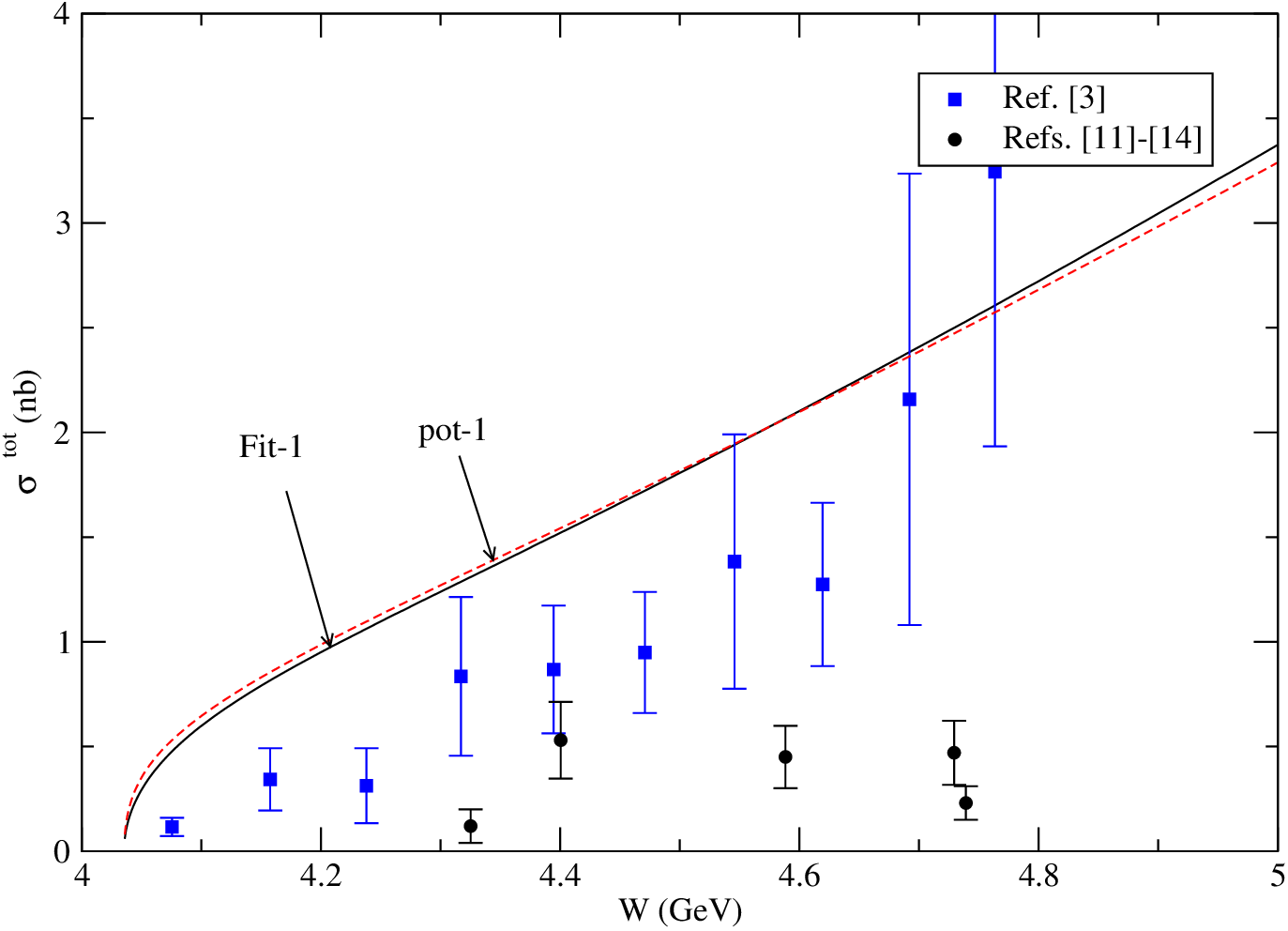}
\caption{Left: Fits to the LQCD data of $J/\Psi$-N potential  of Ref.\cite{sasaki,sasaki-1}.
$v_0$, $\alpha$, and $\beta$ are the parameters of the potential Eq.(\ref{eq:v-lqcd-1}):
$v_0=-0.2$, $\alpha=0.9$ GeV, and  $\beta=1.8$ GeV for Fit-1.
pot-1 is from Fig.\ref{fig:totcrst-sasaki-lqcd}.
 Right: the predicted
total cross sections of  $\gamma +p \rightarrow J/\Psi + p$   are compared
with the data. $F^{off}$ for the VMD parameter $1/f_V$ is set to 1 in these calculations. 
}

\label{fig:totcrst-fit1}
\end{center}
\end{figure}

An another uncertainty of Pomeron-LQCD model is the use of the Yukawa form of Eq.(\ref{eq:v-lqcd}) to fit the 
LQCD data. 
To see how much our results depend on this choice, we 
 now consider  potentials of the following form
\begin{eqnarray}
v(r)= v_0 (\frac{e^{-\alpha r}}{r}-\frac{e^{\beta r}}{r})
\label{eq:v-lqcd-1}
\end{eqnarray}
It differs from Eq.(\ref{eq:v-lqcd}) in having a finite depth at $r=0$:
$v_0(0)=v_0\times(\beta-\alpha)$.
We consider the potentials with $v_0=-0.2$ and $\alpha=0.9$ GeV.
As shown in  the left side of
Fig.\ref{fig:totcrst-fit1}, a potential (Fit-1)  with $\beta=1.8$ GeV can
 reproduce the LQCD data as good as pot-1, except that they have very different shapes  in the  $r < 0.2$ fm
region.
In the right side, we see that the predicted cross sections from these two models
are almost indistinguishable. This suggests  that in the near threshold region, the predicted cross sections
are mainly determined by the potential at $r > $ about 0.2 fm.
Thus a LQCD calculation which is  accurate for determining the potential down to
$r\sim 0.2$ fm will be sufficient for refining the Pomeron-LQCD model.

\begin{figure}[t] \vspace{+2.cm}
\begin{center}
\includegraphics[clip,width=0.4\textwidth,angle=0]{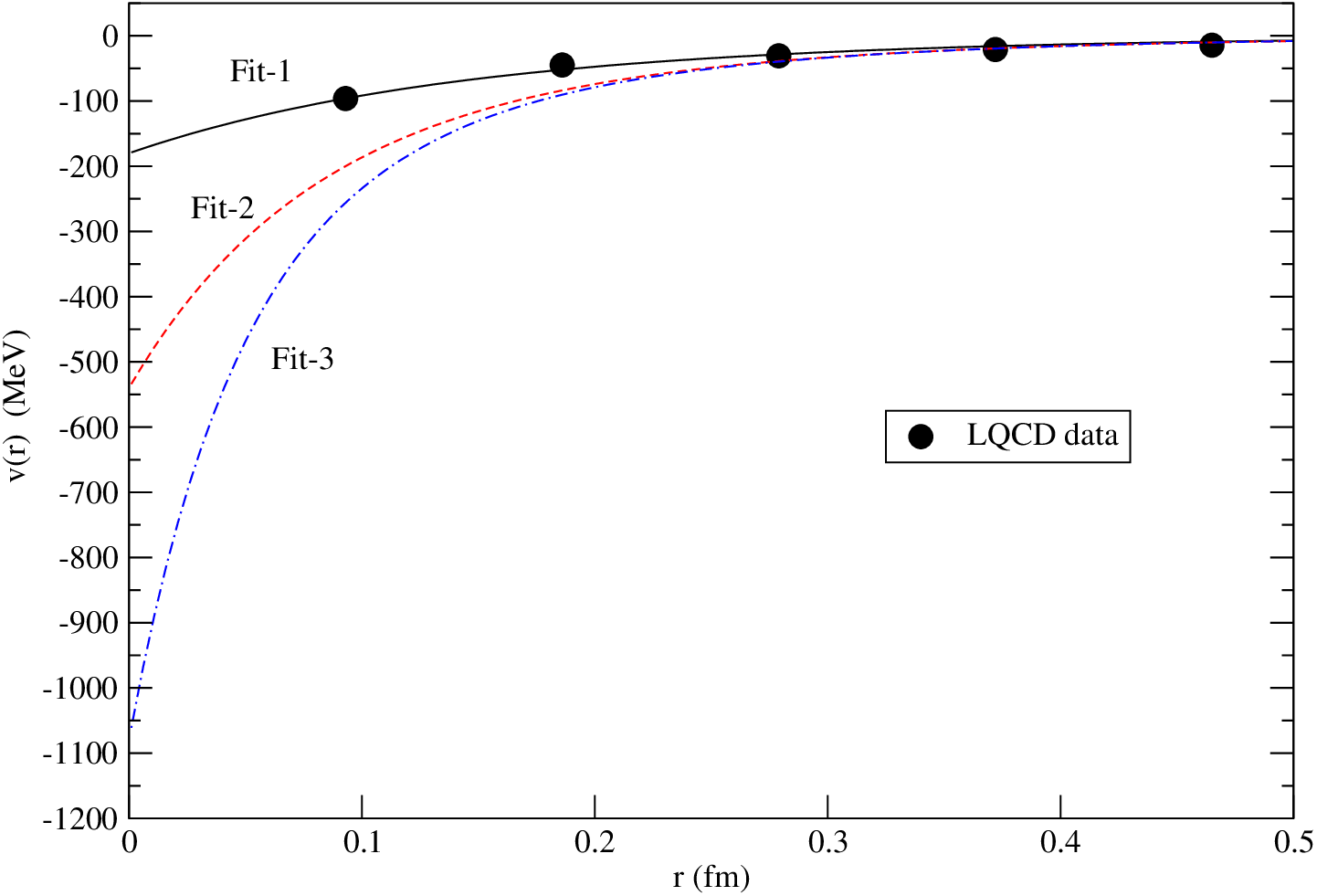}
\includegraphics[clip,width=0.4\textwidth,angle=0]{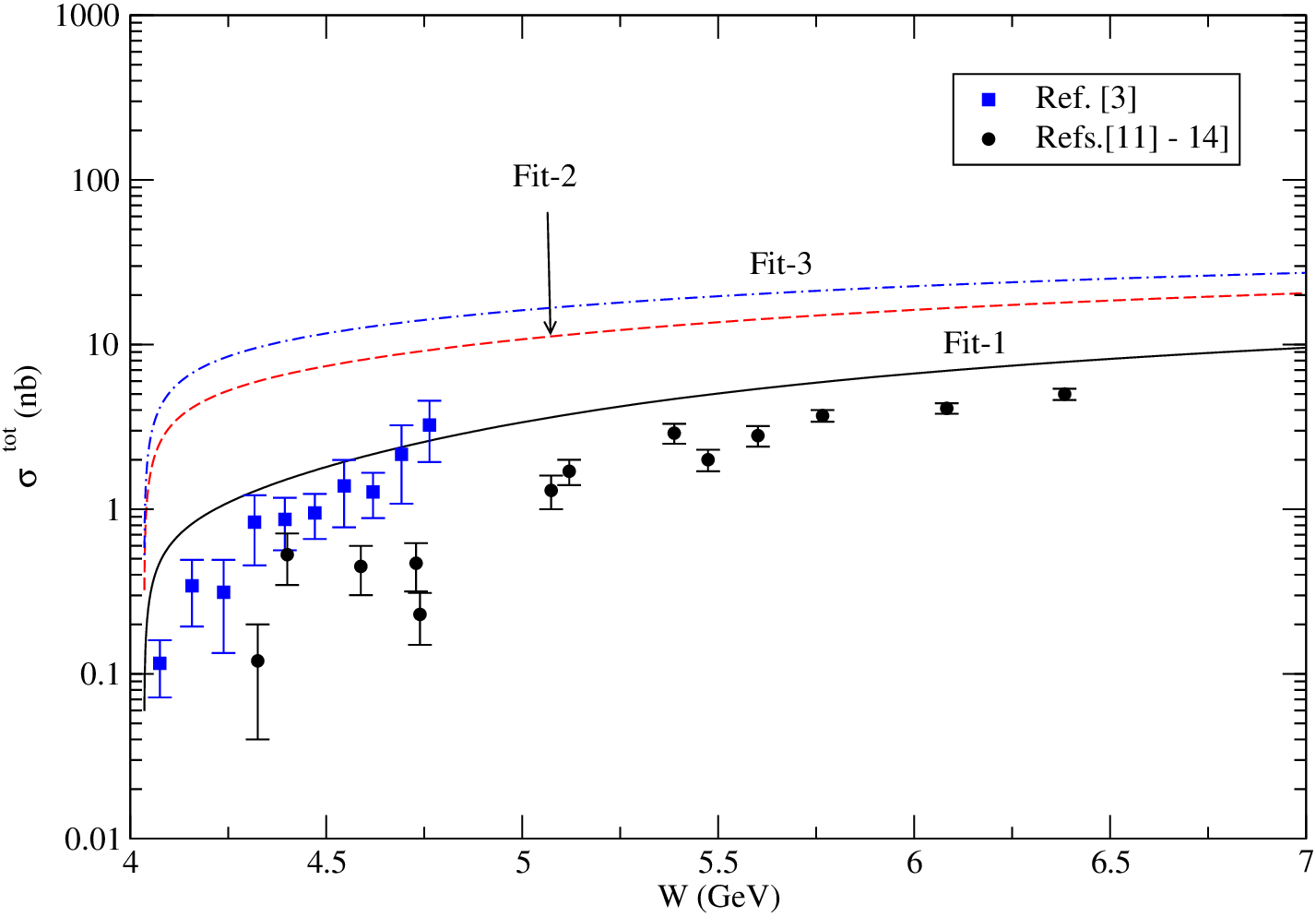}
\caption{Left: Fits to the LQCD data of $J/\Psi$-N potential  of Ref.\cite{sasaki}.
$v_0$, $\alpha$, and $\beta$ are the parameters of the potential Eq.(\ref{eq:v-lqcd-1}):
$v_0=-0.2$, $\alpha=0.9$ GeV and $\beta=1.8, 3.6, 6.3 $ GeV for Fit-1, Fit-2, and Fit-3, respectively.
; Right: the predicted
total cross sections of  $\gamma +p \rightarrow J/\Psi + p$   are compared
with the data. $F^{off}$ for the VMD parameter $1/f_V$ is set to 1 in these calculations.
}

\label{fig:totcrst-fit}
\end{center}
\end{figure}

We next consider two more attractive potentials illustrated in the left side of Fig.\ref{fig:totcrst-fit}.
Their differences with Fit-1  come from using a larger value of $\beta$ in 
Eq.(\ref{eq:v-lqcd-1}) : $\beta = 3.6, 6.3 $ GeV for Fit-2 and Fit-3, respectively. 
We see that  Fit-2 and Fit-3 only fit the LQCD data at $r > 0.3$, and
these three potentials have rather different magnitudes at  $r=0$.
 With $F^{off}=1$, the corresponding predictions of $\sigma^{tot}$ are compared
in the right side of the same figure. It is clear that the magnitudes of the predicted
cross sections 
increase as the  potential becomes more attractive at short distances. Consequently, a smaller $F^{off}$ is needed
to fit the data; $F^{off}= 0.75, 0.4, 0.3$ for Fit-1, Fit-2, Fit-3, respectively.
This is similar to what we have observed using the
potential with Yukawa form Eq.(\ref{eq:v-lqcd}) which approaches  $\infty$ as $r\rightarrow 0$.
Thus the dependence of $F^{off}$ on the attraction of potential is rather independent of
the parametrization of the potential in fitting the LQCD data at $r > 0.3$ fm. 

In summary, we have constructed a  Pomeron-LQCD model of $J/\Psi$ photo-production on the nucleon.
It is based on 
 the $J/\Psi$ potential extracted from a LQCD calculation\cite{sasaki}  and 
the amplitudes generated from the Pomeron-exchange model developed in
Refs.\cite{dl,ohlee,wulee}.
The predicted cross sections are comparable to the recent JLab data at low energies and can also
describe the available data up to $W=300$ GeV. However, a off-shell factor $F^{off}$ for accounting for
the $q^2$-dependence of the VMD constant $1/f_{J/\Psi}$ must be included to explain the JLab data.
It is found that this off-shell factor $F^{off}$ sensitively depends on the short-range part of
$v_{LQCD}(r)$ at $r <$ about  0.4 fm.
To reduce the  uncertainties of the Pomeron-LQCD model
constructed in this work, we  not only
need to have information from  current LQCD calculations  to verify or improve  the $v_{LQCD}(r)$ from
Ref.\cite{sasaki}, but also need to
find a way to predict $F^{off}$ from a QCD model, such as the $q\bar{q}$-loop model explored in Ref.\cite{pich}.

\clearpage
\begin{acknowledgments}
I would like to thank Shoichi Sasaki for providing the information on the $J/\Psi$-N potentials from LQCD
of Ref.\cite{sasaki}.
This work is supported by
the U.S. Department of Energy, Office of Science, Office of Nuclear Physics, Contract No. DE-AC02-06CH11357.
\end{acknowledgments}

\end{document}